\documentclass[pra,aps,twocolumn,superscriptaddress]{revtex4-1}

\usepackage{graphicx}
\usepackage{amsmath}
\usepackage{amssymb}
\usepackage{mathrsfs}
\usepackage{tabularx}

\newcommand{\bs}[1]{\boldsymbol{#1}}

\newcommand{\bckt}[1]{\left[ #1 \right]}
\newcommand{\abs}[1]{\left| #1 \right|}

\newcommand{\mrm}[1]{{\mathrm{#1}}}

\newcommand{\meq}[1]{\begin{equation} #1 \end{equation}}
\newcommand{\mea}[1]{\begin{align} #1 \end{align}}
\newcommand{\val}[2]{\ensuremath{#1 \; \mathrm{#2}}}	

\begin{document}

\title{Perturbation-induced defects in trapped superfluids exhibit generic
behavior}

\author{Peter Scherpelz}
\author{Karmela Padavi\'c}
\author{Andy Murray}
\affiliation{James Franck Institute and Department of Physics,\\ University of Chicago, Chicago, Illinois 60637, USA}
\author{Andreas Glatz}
\affiliation{Materials Science Division, Argonne National Laboratory,\\ 9700 South Cass Avenue, Argonne, Illinois 60439, USA}
\affiliation{Department of Physics, Northern Illinois University, DeKalb, Illinois 60115, USA}
\author{Igor S. Aranson}
\affiliation{Materials Science Division, Argonne National Laboratory,\\ 9700 South Cass Avenue, Argonne, Illinois 60439, USA}
\author{K. Levin}
\affiliation{James Franck Institute and Department of Physics,\\ University of Chicago, Chicago, Illinois 60637, USA}


\begin{abstract}
We investigate equilibration processes
shortly after sudden perturbations
are applied to
ultracold trapped superfluids.
We show the similarity of
phase imprinting and localized
density depletion perturbations, both of which initially are found to
produce ``phase walls". These planar defects are
associated with
a sharp gradient in the phase. Importantly
they
relax following a quite general sequence.
Our studies, based on simulations of the complex time-dependent Ginzburg-Landau
equation,
address the challenge posed by these experiments:
how a
superfluid eventually eliminates a spatially extended planar defect.
The processes involved are necessarily more complex than equilibration
involving simpler line vortices.
An essential mechanism for relaxation involves repeated formation and
loss of vortex rings near the trap edge.
\end{abstract}

\maketitle

\section{Introduction}

Our understanding of superfluid equilibration has been greatly
enhanced by research in trapped ultracold atomic gases.
In contrast to condensed matter systems, here there is flexibility in 
engineering 
a highly controlled perturbation, as well as the capability of
studying the ensuing relaxation processes in real time
experiments. 
Such studies inform about the nature of
non-equilibrium superfluid dynamics and equilibration
processes.
Of particular interest are the nature of 
defects formed by perturbations, and the life cycle of these defects as they
equilibrate.
In this paper we focus on extended planar defects in two and three dimensions
and establish
(from both numerical observation and analytical
arguments) the processes whereby these evolve during equilibration. 
We point out that these planar defects which we call
``phase walls" (as in ``domain walls"-- surfaces
along which there is a sharp gradient of the phase)
are rather ubiquitous; they undergo a
reproducible evolutionary sequence towards
a simpler defect, as for example a single line vortex. 
In contrast to line vortices which tend to precess within the
trap before exiting, the healing processes of these more extended defects 
are quite complex. 

Indeed, it is not obvious how
even a commonly-studied planar defect, a dark soliton
(stable in one dimensional systems) eventually decays in realistic 2D or 3D
traps, as it must.
What we show here is that the decay of extended planar defects
involves vortex rings. But these rings are not the end-product
of a process in which the plane disappears, nor are they
in general associated with a 
``soliton-ring oscillation" process \cite{shomroni_2009}. Rather
there is a co-existence of phase wall with
near-trap-edge vortex rings which repeatedly
enter and exit the trap.

Our work is based on simulations and analysis of
two types of experiments which give rise to planar defects. These include applying
localized
depletions of density or alternatively subjecting the system to phase-imprinting,
where half the gas
is subjected to a phase change $\Delta \Phi$.
We focus here on
a more asymmetric phase imprint with
$\Delta\Phi = 130^\circ$. This configuration is less well-studied
experimentally, but it is of interest to us here
because it creates faster-moving defects
in the fluid, and more clearly reveals the full time evolution
of the equilibration process.

Importantly, in this paper we demonstrate
the generality of the induced defect structures
and their dynamics within these two different perturbations.
Our work establishes predictions for future experiments; we
address accessible time scales (in the ms range) and in
our simulations we consider
trapped gas parameters which match experimental capabilities.
Finally, our work
facilitates the more general understanding of the dominant
equilibration pathways present in trapped atomic superfluids when they are
perturbed.

In addition to suggesting directions for future experimental
research, these
studies yield
further insight into related past experiments and simulations
\cite{andrews_1997,ruostekoski_2005,shomroni_2009,becker_2013,yefsah_2013,
ku_2014}.
In the process we
clarify the many different defect sequences which have
been reported in the literature.
Indeed, there has been significant controversy 
and debate over the characterization
and decay processes of various defects, notably those associated with
a phase imprint in which 
$\Delta \Phi 
\sim \pi$
where the original defect was thought to be a heavy solition 
\cite{yefsah_2013}, 
and later a vortex ring \cite{bulgac_2014} and still later, as was
consistent with predictions from some of our co-authors \cite{scherpelz_2014}, 
established
to be a solitonic vortex \cite{ku_2014}.
Similar mis-identifications were associated with a different class of
perturbations 
\cite{lamporesi_2013,donadello_2014}.
It should be noted that while this earlier work \cite{scherpelz_2014} with $\Delta \Phi \sim \pi$
is a special case leading to anomalously long-lived defects,
nevertheless, the earlier stages of equilibration (approximately the first
5 ms) in this case
too, are addressed by the defect
sequence we report here.

Among the earliest controlled perturbation experiments
in cold gases 
\cite{andrews_1997} involved establishing propagating, localized depletions of
density through a focused laser beam. While these were designed to measure the
speed of sound, these studies very early on elicited somewhat
controversial \cite{stamper-kurn_1999} theoretical suggestions
that dark solitons (planar density depletions, stable in 1D and
associated with a phase shift)
had been created in the process
\cite{reinhardt_1997,jackson_1998,hong_1998,
brazhnyi_2003,ruostekoski_2005}. Similar experiments have led to
the observation of a variety of defects
\cite{dutton_2001,ginsberg_2005,weller_2008,shomroni_2009}, while
other studies engineered localized excitations through 
phase imprinting
\cite{burger_1999,denschlag_2000,becker_2008,becker_2013,yefsah_2013,ku_2014}.
The theory associated with these experiments has suggested that solitons, or
vortices or vortex rings (in some combination or sequentially) are created in
the process
\cite{becker_2013,cetoli_2013,wen_2013,reichl_2013,bulgac_2014,
scherpelz_2014,ku_2014,wlazlowski_2014}.

\begin{figure*}
\includegraphics[scale=0.92]{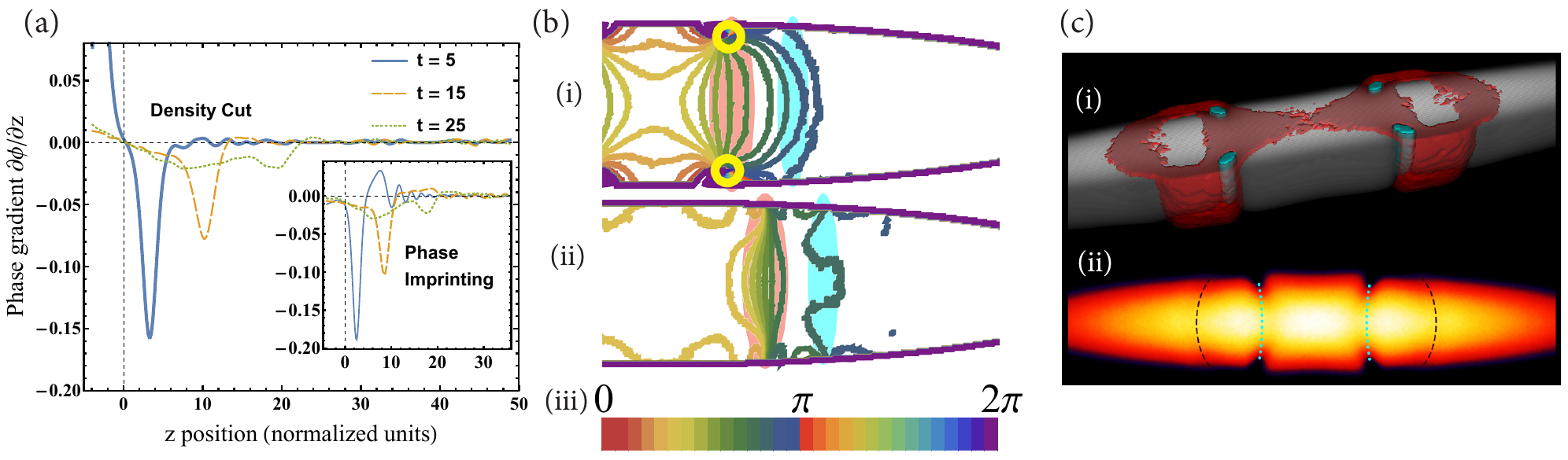}
\caption{(Color online) 
Figures illustrating the coexistence of a vortex ring and phase wall, along with
the subsequent ``tearing'' process, in 3D simulations. 
(a) A plot of the gradient of the phase along the long axis ($z$-axis with
$x=y=0$), with data for both density cut and phase
imprinting simulations at early times. 
It shows that in both density cut and phase imprinting
perturbations, a sharp phase wall forms ($t = 5$). This is
followed by the phase wall spreading out ($t=15$) and finally by separating
into a front phase wall and a vortex pair behind it, as shown by the two dips
at $t=25$.
(b) 
Phase contour plots of $x=0$ planes for two frames of the
density cut simulation in (a), demonstrating the tearing process that occurs
subsequent to the phase wall separation in (a). 
(b.i) Frame $t = 25$, which displays the system as the phase wall and vortex
ring are 
``bending apart''. 
(b.ii) Frame $t = 39$, displaying the
system after tearing has occurred.  Two separate phase walls now exist, one 
due to the original phase wall and the other due to the vortex ring.
In this specific simulation, no vortices are present at
the moment, but two are re-nucleated soon after this point. 
As shown in the legend (b.iii), $\pi/15$ contours of the
phase are displayed.  Yellow circles, red ellipses, and cyan ellipses highlight
vortices, the vortex ring defect structure, 
and the original phase wall respectively.
(c) Visualizations of the tearing process itself, showing the 
coexisting vortex ring and phase wall at $t=32$ in the 3D density depletion
simulation from (a) and (b). (c.i) 3D visualization of half the cloud 
(cut along the $y=0$ plane).  The background (white) is an isosurface of
$\abs{\Psi}^2$, which provides an outline of the cloud. The diffuse red surface 
illustrates the region of moderate volume depletion from the equilibrium cloud,
and the cyan curves indicate the vortex rings (large
volume depletion). (c.ii) Amplitude in the $x=0$ plane
from $\abs{\Psi}=0$ (black) to $\abs{\Psi}=\abs{\Psi_\mrm{max}}$ (white).
The dashed curves mark the front phase wall causing a density depletion, while
the dotted cyan curves mark the location of the vortex ring. 
In all of (a-c), the ``density cut'' shows data after removing at $t = 0$ a 
Gaussian density depletion $V = Ae^{-z^2/(2 \sigma^2)}$ with $A = 3.6$
and $\sigma =  0.71$. The phase imprinting shows data following
a $130^\circ$ phase imprinting at $t=0$. 
\label{fig:tearing}}
\end{figure*}

The challenge presented by these perturbations has to do with how the
superfluid system eventually eliminates or heals this planar defect as
equilibration progresses.
Quite generally we find that equilibration involves two types of
processes. First, some equilibration processes will
progressively extend the
phase change $\Delta\phi$ over a larger volume, decreasing the maximum 
magnitude of the local phase gradient. Others will instead diminish the value of
$\Delta\phi$ itself, minimizing the phase difference 
between the two sides.  Facilitating these healing
processes are 
vortex rings which repeatedly
enter and exit the trap, and without which the overall sequence of equilibration
cannot occur.
We also report that the phase walls, which lead to vortex ring
nucleation, tend to persist with (rather than convert to) vortex rings.
In this paper we will show that
all of these processes, which are to be associated with the planar nature of the
initial defect, are in fact rather general.

In contrast to the commonalities
upon which we focus here, past literature
has suggested a wide variety of different defect evolutions.
These include the report of a so-called oscillating soliton--vortex ring
\cite{shomroni_2009} created with a density depletion, while in a more
complicated system of two density defects, there are claims of
structures that involved both solitons and vortex
rings \cite{ginsberg_2005}.
Both phase imprinting and density depletions have led to
reports of multiple solitons that
bent significantly and in some cases decayed to vortices
\cite{denschlag_2000,dutton_2001}.
Finally, numerical
simulations of phase imprinting in very anisotropic traps led to the
identification of a very weak
vortex ring which rapidly moved away from the trap center
\cite{becker_2013}. Here we show how many of these defects are related, while a
few arise due to different limits of trap geometry.

We focus on cigar-shaped traps which mirror those
used in recent experiments \cite{shomroni_2009,ku_2014,donadello_2014}. 
These 
elongated traps, which are well into the 2D or 3D regime, provide the advantage
that one can study early-time dynamics without encountering reflections from
trap edges. Furthermore, the cigar shape leads to significant and 
realistic inhomogeneity in the radial direction, which
strongly influences defect behavior in these experimental systems.
We will demonstrate
that this inhomogeneity is, in fact, central for understanding most of the
behavior seen here. 

Our analysis is based on numerical simulations of the complex time-dependent
Ginzburg-Landau (TDGL) equation for Fermi gases \cite{sademelo_1993}. 
These studies address only the condensate dynamics and do not
include the effects of fermionic excitations. Considerable support
for these numerics comes from previous related studies 
\cite{scherpelz_2014}
of phase imprinting 
near $\Delta \Phi = \pi$ where most aspects of the 
experiments 
\cite{ku_2014} confirmed the earlier predictions.
In
such simulations it is important to avoid artificially
symmetric situations which unphysically stabilize long-lived defects. Here
we include
stochastic noise. The equation one solves is very similar
to the Gross-Pitaevskii (GP) equation with the inclusion of dissipation. 
Thus, for the most part, our low dissipation results apply to the
condensate dynamics in
Bose gases as well.

\section{Our approach} 
Our calculations are based on simulations of the wavefunction dynamics of a
condensate in the Bardeen-Cooper-Schrieffer (BCS) to Bose-Einstein condensate
(BEC) crossover, using an extension of TDGL to this crossover
originally developed in Ref.~\cite{sademelo_1993}. 
Following Ref.~\cite{abrahams_1966} these authors derived the
dynamical equation for the order parameter or gap $\Delta$ in the form 
\begin{equation}
 \bckt{ a + b |\Delta(\bs{ x}, t)|^2 - \frac{c}{2m} \nabla^2 - i d
\frac{\partial}{\partial t} } \Delta (\bs {x} , t) = 0
\label{eq:tdgl1}
\end{equation}
We work with the TDGL equation in a rescaled form \cite{aranson_2002},
\mea{
    e^{-i\theta} & \partial_{\hat t}\hat \Psi(\hat{\bs{x}},\hat t) =   \label{eq:TDGL} \\
    & \{[1-\hat V(\hat{\bs{x}})]+\frac{1}{2}\hat\nabla^2-|\hat\Psi(\hat{\bs{x}},\hat 
t)|^2\}\hat\Psi(\hat{\bs{x}},\hat t)+\chi(\hat{\bs{x}},\hat t). \notag
}

Time $t$, position $\bs x$, order parameter $\Delta$ and potential $V$ 
have been converted to dimensionless normalized quantities,
\meq{
\hat t = \frac{-a}{\abs{d}}t, 
\quad \hat{\bs x} = \sqrt{\frac{-ma}{c}}\bs x, \quad 
\hat \psi = \sqrt{\frac{b}{-a}}\Delta, \quad \hat V = \frac{a_1}{-a}V.} 
With this, Eq.~\eqref{eq:tdgl1} reduces to Eq.~\eqref{eq:TDGL}, with 
$\theta = \pi/2 - \arg(d)$ and stochastic noise $\chi$ (discussed below) added.

Here $\theta$ controls the amount of dissipation, allowing us to move
from the BCS ($\theta=0$) regime to the BEC regime; for $\theta=\pi/2$ 
Eq.~\eqref{eq:TDGL} 
reduces to the time-dependent GP equation \cite{sademelo_1993,pethick_2008}. 
Work by one of the present authors shows that the inclusion 
of a gap in the fermionic spectrum stabilizes the pairs, making them rather
long-lived \cite{maly_1999}.
As a result, we typically use $\theta=88^{\circ}$ in our simulations, and thus
expect our results to also apply directly to BEC condensates 
\footnote{We see little
variation in results for simulations with lower dissipation, $\theta =
89.5^\circ$. Other dissipation effects mirror those shown in the Supplemental
Material of Ref.~\cite{scherpelz_2014}.}.

 In the weak
coupling limit, sample parameters of $T_c = \val{70}{nK}$, $T/T_c = 0.7$, and
coherence length $\xi = \val{3.2}{\mu m}$ can be used with
Ref.~\cite{sademelo_1993} to give the 
unit of $\hat t$ as $\val{0.12}{ms}$ and the unit of $\hat x$ as $\val{1.0}{\mu
m}$.
The rescaled equation can similarly be obtained from the Gross-Pitaevskii
equation \cite{pethick_2008} using the analogous transformations above. In this
case, using pairs of ${}^6$Li atoms with a chemical potential $\mu =
\val{120}{nK}$ (approximately equal to that reported in Ref.~\cite{yefsah_2013})
is sufficient to set the physical time and length scales, namely 
a unit of $\hat t$ as $\val{0.06}{ms}$ and a unit of $\hat x$ as $\val{0.6}{\mu
m}$. More information can be found in the Supplementary Material of 
Ref.~\cite{scherpelz_2014}.

In this work we consider
2D and 3D anisotropic trapped Fermi gases subjected to 
either phase or density perturbations. 
The trap potential $\hat V$ in Eq.~\eqref{eq:TDGL} is inserted by using the 
local density approximation, $\mu \rightarrow \mu - V(\bs{x})$, with the 
chemical potential $\mu$ rescaled to unity, so that 
$\hat V(\bs{x})=\hbar(\omega_{\perp}^2(x^2+y^2)+\omega_z z^2)/(2\mu)$.
The function $\chi$ represents uniformly distributed thermal noise, and a 
fluctuation temperature can be set through $\langle 
\chi(\bs{x},t)\chi^*(\bs{x}',t')\rangle=2
T_{\chi}\delta(\bs{x}-\bs{x}')\delta({t-t'})$. 
We use a dimensionless fluctuation $T_\chi$ temperature very close 
to the zero-temperature limit (well below a nanokelvin for cold gas systems)  
\footnote{The temperature corresponds to $T_{\chi}=5\times10^{-9}$. Note that
this parameter includes a damping term $\gamma_\chi$ 
from atomic collisions, $T_{\chi} = \gamma_\chi
T_{\mrm{cloud}}$, which we approximate as $10^{-2}$ and is
discussed elsewhere \cite{damski_2010}.}.
We stress that the 
inclusion of noise in our numerical approach is a significant addition
\cite{scherpelz_2014} that avoids 
unphysical cylindrically-symmetric systems.
Many 
other recent studies concerned with solitons and vortex
structures in ultracold quantum gases
\cite{ruostekoski_2005,reichl_2013,bulgac_2014,wlazlowski_2014} omit noise and
other symmetry-breaking mechanisms.

Our work is based on numerical simulations discretized in up to $8192\times1024$
(2D) or $1024\times256^2$ (3D) grid 
points and designed to solve Eq.~\eqref{eq:TDGL} using a quasi-spectral 
split-step method.

\section{Superfluid perturbations}
For our simulations, 
we study a cigar-shaped trap with axial symmetry, typically taking
$\gamma = \mu/(\hbar\omega_\perp) =8.0$ and a trap ratio 
$\lambda = \omega_{\perp}/\omega_z = 6.6$ unless otherwise noted. (This
corresponds to radial trapping frequences of 160 Hz and 320 Hz 
for the two example physical systems above.)
In this regime the
system is strongly under the influence of trap inhomogeneity, but well
outside the 1D limit. Similar results for inhomogeneous,
cylindrical traps are presented in App.~\ref{sec:cyl} below.
In the density depletion or ``density cut" perturbation 
\cite{andrews_1997, shomroni_2009}
the superfluid equilibrates in the presence of a 
sharp barrier located about the $z=0$ plane, physically corresponding to the
application of a blue-detuned laser. This density cut is then removed
suddenly, allowing the two sides of the cloud to interact.

Alternatively, we apply a ``phase imprint'' in which
half of the cloud's phase is rotated by an angle $\Delta\Phi$.
In contrast to recent experiments 
\cite{ku_2014,yefsah_2013}
in which
$\Delta\Phi \approx 180^\circ$
was considered, here
we focus on a less symmetric situation in which the phase
difference is
$\Delta\Phi = 130^\circ$ and the defects evolve more rapidly.
We note that the phase imprint technique has
been frequently used to create solitary waves in superfluids
\cite{burger_1999,denschlag_2000,becker_2008,becker_2013,yefsah_2013,ku_2014}.

\section{Defect overview}
Figure \ref{fig:tearing} summarizes the major effects that
we see in a 3D simulation. 
The general features we observe begin with the creation of a sharp phase
gradient, either directly in phase imprinting or, as in a density depletion
perturbation, through rapid movement of the
superfluid. This ``phase wall'' moves and spreads out, then
bends outward in its center, soon nucleating a vortex ring on the boundary. This
process can be seen in Fig.~\ref{fig:tearing}(a), where one dip in the
phase gradient splits into two dips at later times, one due to
the front phase wall and the other due to the vortex ring.

Importantly, the front phase wall persists and bends forward after 
nucleation of the vortex ring. 
Eventually the phase wall and the vortex ring ``tear'' from each
other, separating completely. This process can be seen in
Fig.~\ref{fig:tearing}(b-c), where it is also accompanied by the ejection (and
later, re-nucleation) of vortex rings.

These processes, as will be shown in subsequent simulations, are quite general.
The bending of the phase wall causes vortex nucleation, and 
once vortices are nucleated, the ``tearing'' processes cause the initial 
phase wall
to break up into multiple defects. We will next focus on more detailed analyses
of these phenomena in 2D.

\begin{figure}
    \includegraphics[scale=0.7]{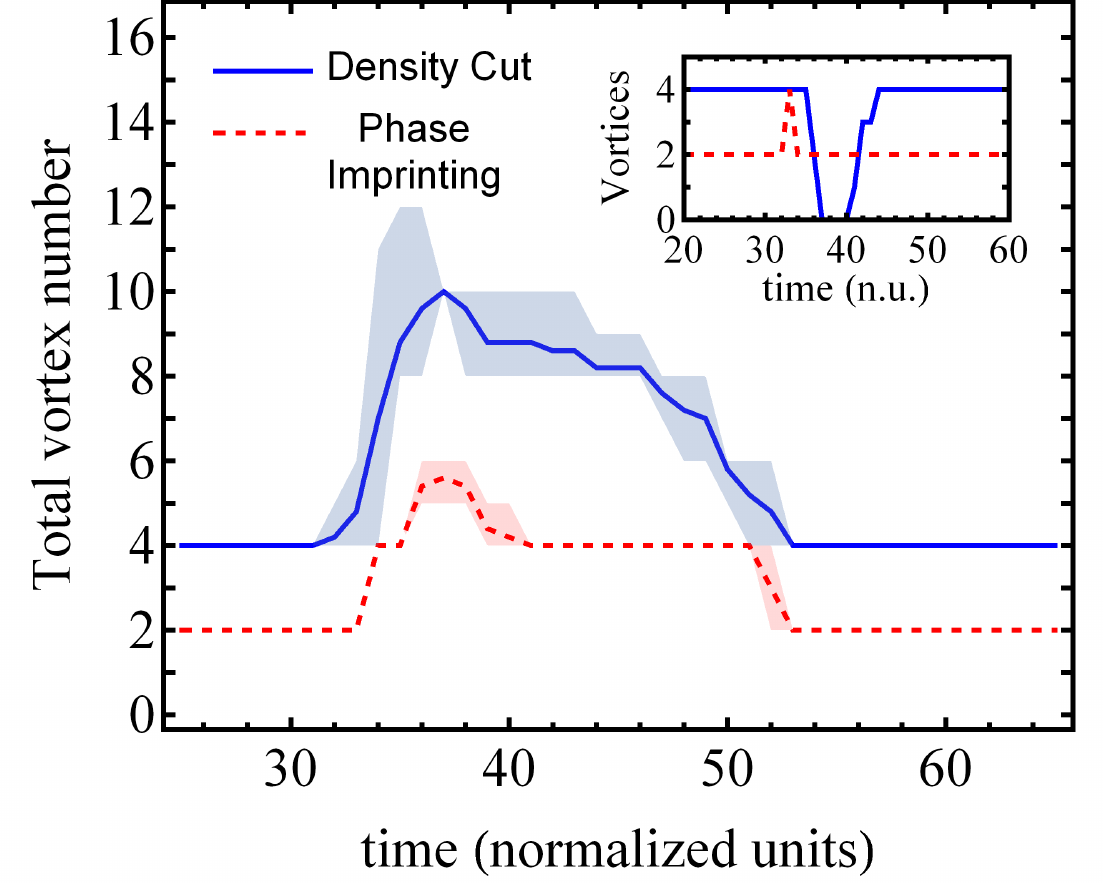}
    \caption{
Plots of the total vortex number (irrespective of vortex orientation) 
vs.\ time for 2D simulations. 
Five runs with unique random noise were used for each perturbation. 
The curves show the mean vortex number, 
while the shaded regions about the curves show the full range of
vortex numbers found at each timestep. 
This plot shows that density cuts and
phase imprinting plots follow the same tearing process, first forming a
vortex/antivortex pair (jump in vortex number near $t = 34$) 
and then ejecting the antivortices near $t=38$. Finally, one of the remaining
vortex pairs is ejected. Because density cuts produce vortices in both halves of
the trap, all vortex numbers are doubled for it compared to phase imprinting.
The inset shows vortex number for the two 3D runs in Fig.~\ref{fig:tearing}.
Both exhibit a tearing process around $t = 35$, but because the vortex rings are
close to the trap edge, the order of vortex ejection and nucleation differs
somewhat -- for a density cut, the original vortex ring exited 
before a new vortex ring entered, while for phase imprinting, the anti-vortices
formed too close to the trap edge to be resolved in the measurements here.
}
    \label{fig:vnum}
\end{figure}

\begin{figure*}
\includegraphics[scale=0.9]{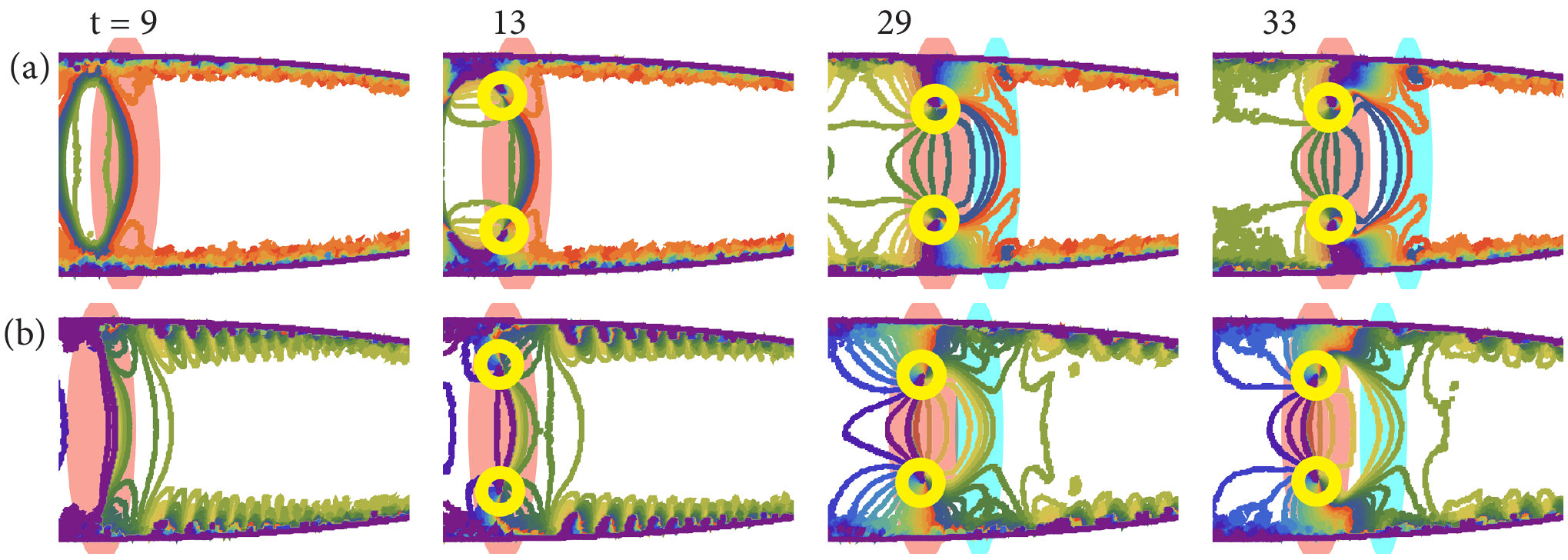}
\caption{Phase contour plots of significant frames for the 2D (a) density cut
    and (b) phase imprinting runs, matching the parameters used for the
    simulations in Fig.~\ref{fig:tearing}, and 
    using the same contours as Fig.~\ref{fig:tearing}(b). Red highlights the
    phase wall closer to the center, cyan the phase wall that will ``tear'',
    and vortices are circled in yellow.  
    The first frame at $t = 9$ shows the initial bending of the
    phase wall, followed by the nucleation of the initial vortex pair at
    $t=13$.  The following frames show the beginning of the tearing
    processes, with a phase wall bending out from the vortex pair and its
    accompanying phase gradient. 
\label{fig:contours1}}
\end{figure*}

\begin{figure}
    \includegraphics[scale=0.4]{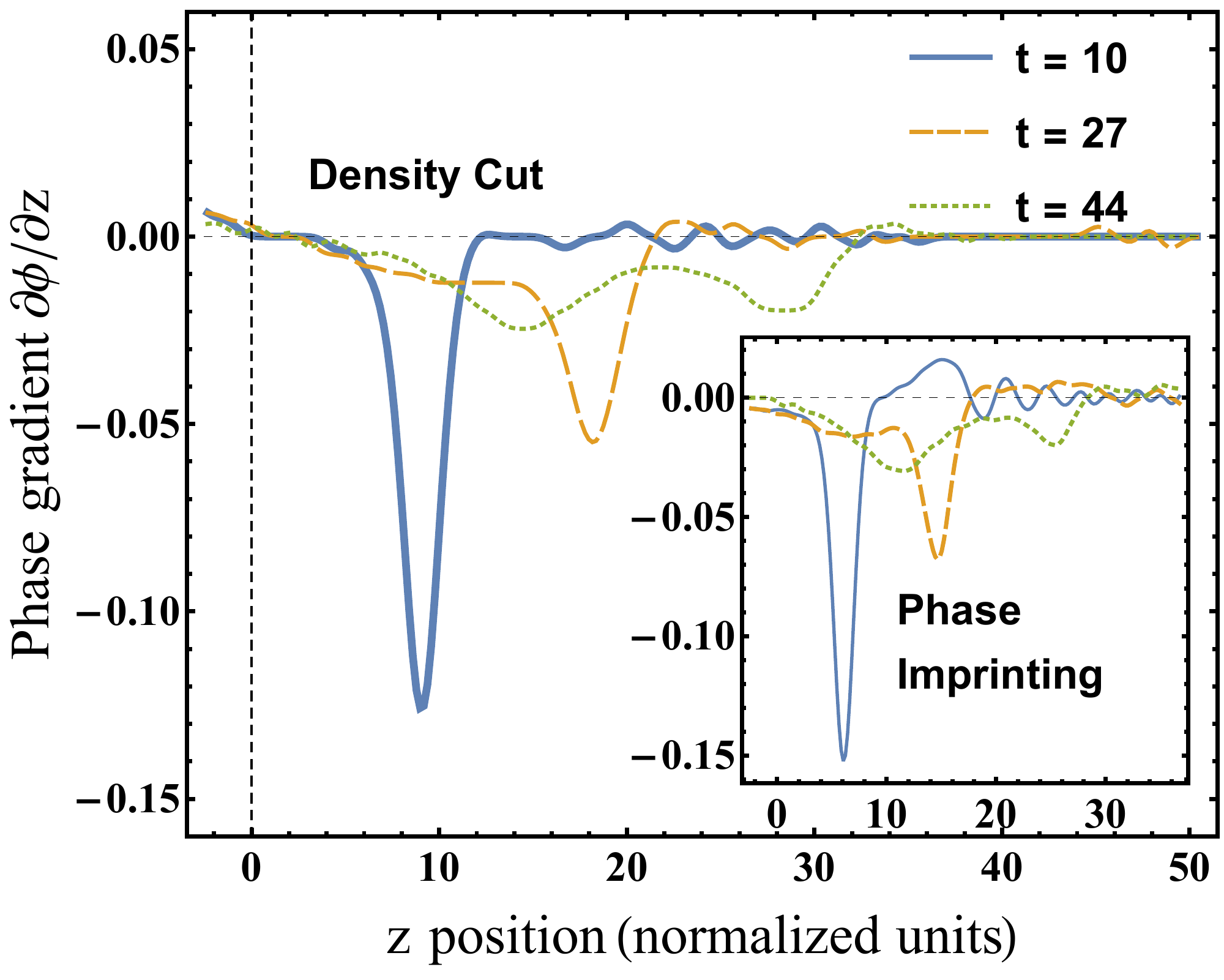}
    \caption{The evolution of the phase gradient along the axis of the trap for
        2D simulations. The phase gradient starts out with a 
        large, sharp dip, which later spreads out and then splits into two as
        the vortex pair and the original phase wall separate. 
        As can be seen by comparing to Fig.~\ref{fig:tearing}(a), the timescales
        differ but the process of evolution matches very well for 2D and 3D
        simulations.
    }
    \label{fig:pgtwod}
\end{figure}

\begin{figure*}
\includegraphics[scale=0.9]{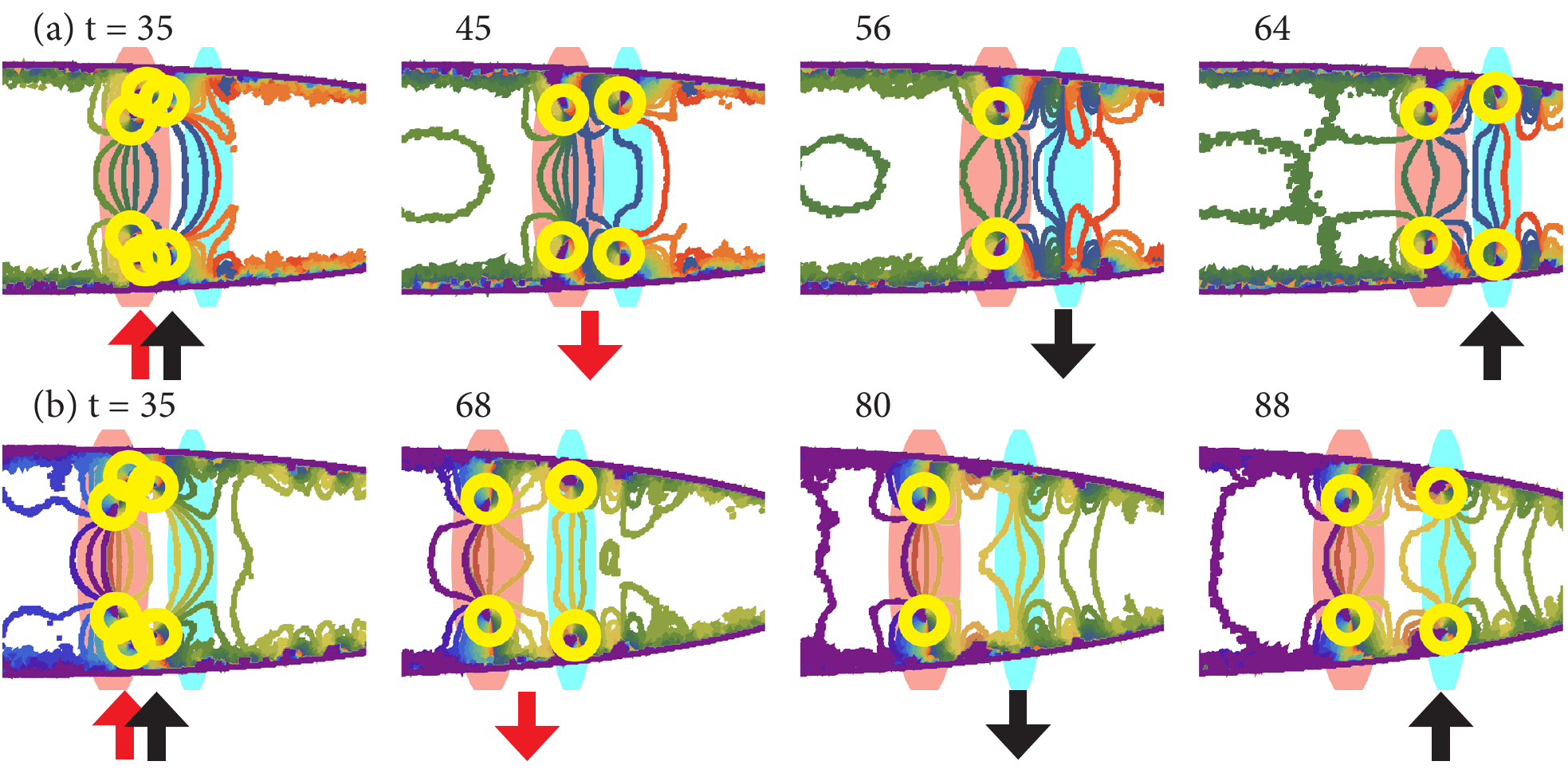}
\caption{Phase contour plots of the simulations in Fig.~\ref{fig:contours1} at
later times. Here black (red) arrows show
vortices (antivortices) entering or leaving the cloud.
The first frame at $t = 35$ shows the conclusion of the tearing process, with
the nucleation of vortex-antivortex pairs and the separation of the two phase
walls. The second frame [$t = 45$ in (a) and $t=68$
in (b)] shows the loss of the antivortex, while the third frame shows the
ejection of a vortex, followed by its renucleation in the final frame. Note
that in (b), the final three frames show a portion of the cloud somewhat closer
to the edge than all other frames.
\label{fig:contours2}}
\end{figure*}

\section{General defect sequence}
Figure \ref{fig:vnum} shows the evolution of vortex number for a sample of 2D
simulations, also compared to the evolution of vortex number in the 3D runs of
Fig.~\ref{fig:tearing}. The 2D simulations are quite consistent internally, and
although the exact vortex number progression varies between 2D and 3D, all runs
show similar perturbations in vortex number due to the tearing processes. In
order to both explore this process more detail, and the generality of these
effects under many different trap parameters, we now turn to a more microscopic
analysis of 2D runs.

Figure \ref{fig:contours1} shows the evolution of the phase in the beginning
moments of these simulations. 
Initially one sees a bending of the phase wall, which is a direct
consequence of trap inhomogeneity. The phase wall is equivalent to a deformed
dark soliton, which moves at a speed dependent on the phase change $\Delta\phi$
across it, as well as the local speed of sound. Due to the large central
density, the speed of sound is highest at $(x,y) = 0$, and thus the
phase wall moves fastest on the central axis \cite{parker_2006,shomroni_2009}. 
In contrast, the negligible density at the edge of the cloud effectively pins
the phase wall on the boundary, establishing the dominant role of radial
inhomogeneity.

This bending leads to the first step in the sequence shown in
Fig.~\ref{fig:contours1} which is
the nucleation of a vortex
pair at the boundaries (in 3D this corresponds to a vortex ring). 
The
bent phase wall causes a strong superfluid flow toward the
cloud center. This flow along the edge nucleates the vortex pair, and also
pulls the vortex pair inward, as shown in the second frame of 
Fig.~\ref{fig:contours1}.
This step has been
observed previously in experiments \cite{shomroni_2009} 
and simulations \cite{ruostekoski_2005}, and 
has been reported to play a crucial role in recent studies
\cite{yefsah_2013,reichl_2013,bulgac_2014,scherpelz_2014,
ku_2014,wlazlowski_2014} \footnote{The instability of solitons to vortex
formation in trapped gases has also been modeled theoretically 
\cite{brand_2002,komineas_2002,komineas_2003}, 
though not necessarily in terms of the dynamic nucleation
beginning at the boundary that is dominant here.
We note that this nucleation at the boundary is related to but distinct from a
true snake instability
\cite{kuznetsov_1988,mamaev_1996,mamaev_1996b,tikhonenko_1996,kivshar_1998,
feder_2000,anderson_2001,cetoli_2013}, 
which depends on transverse fluctuations of the soliton position
and forms vortices within the soliton, not just at the boundary. 
As seen in Fig.~\ref{fig:largegamma}, larger systems where
trap inhomogeneity is less prominent can display both mechanisms of vortex
formation \cite{dutton_2001,ginsberg_2005,ruostekoski_2005,parker_2006}.}.

Once nucleated, the vortices have multiple, dramatic effects on the system.
Trap inhomogeneity and boundary effects \cite{mason_2008}
can be qualitatively associated with image vortices \cite{scherpelz_2014}.
These  cause the vortices to move toward the axial ends of the trap.
The vortex cores themselves now form the radial edges of the phase wall, 
so these edges also now move axially outward
rather than being pinned near the trap center along the axial direction. 
Simultaneously, the vortices also
strongly influence the form of the phase gradient along the axis. The
gradient is spread
out axially while it simultaneously begins to separate the original phase
wall from the vortex pair (see Fig.~\ref{fig:pgtwod}).

\subsection{Separation of phase wall and vortices}

Importantly, these vortices do not destroy the original phase wall, but
rather both defects
coexist.
After vortex nucleation (except in very one dimensional systems), 
the phase wall continues bending outwards, with its ends still formed by 
the vortices.
This phase wall
is necessarily weaker than before the vortex formation, due to the
fact that a part of the initial phase change $\Delta\Phi$ now resides 
in the vortex pair structure itself, as shown in Fig.~\ref{fig:pgtwod}.
However, it can be strong enough that, once the bending is severe, a second
vortex pair nucleation occurs, 
as shown at $t=35$ in Fig.~\ref{fig:contours2} and observed in
Fig.~\ref{fig:vnum} \footnote{A similar result of
vortex ring formation with a persistent soliton was reported
in Ref.~\cite{ginsberg_2005}. Reference \cite{ruostekoski_2005}
also finds a second vortex/antivortex nucleation, but does not observe a second
persistent feature form as we do.}.
This tearing process
allows the phase wall to separate entirely from the initial pair of vortices.
 Often, the persistence of the original
phase wall, and this tearing phenomenon, is robust enough that it occurs 
yet a second time.

\begin{figure}
\includegraphics[scale=0.43]{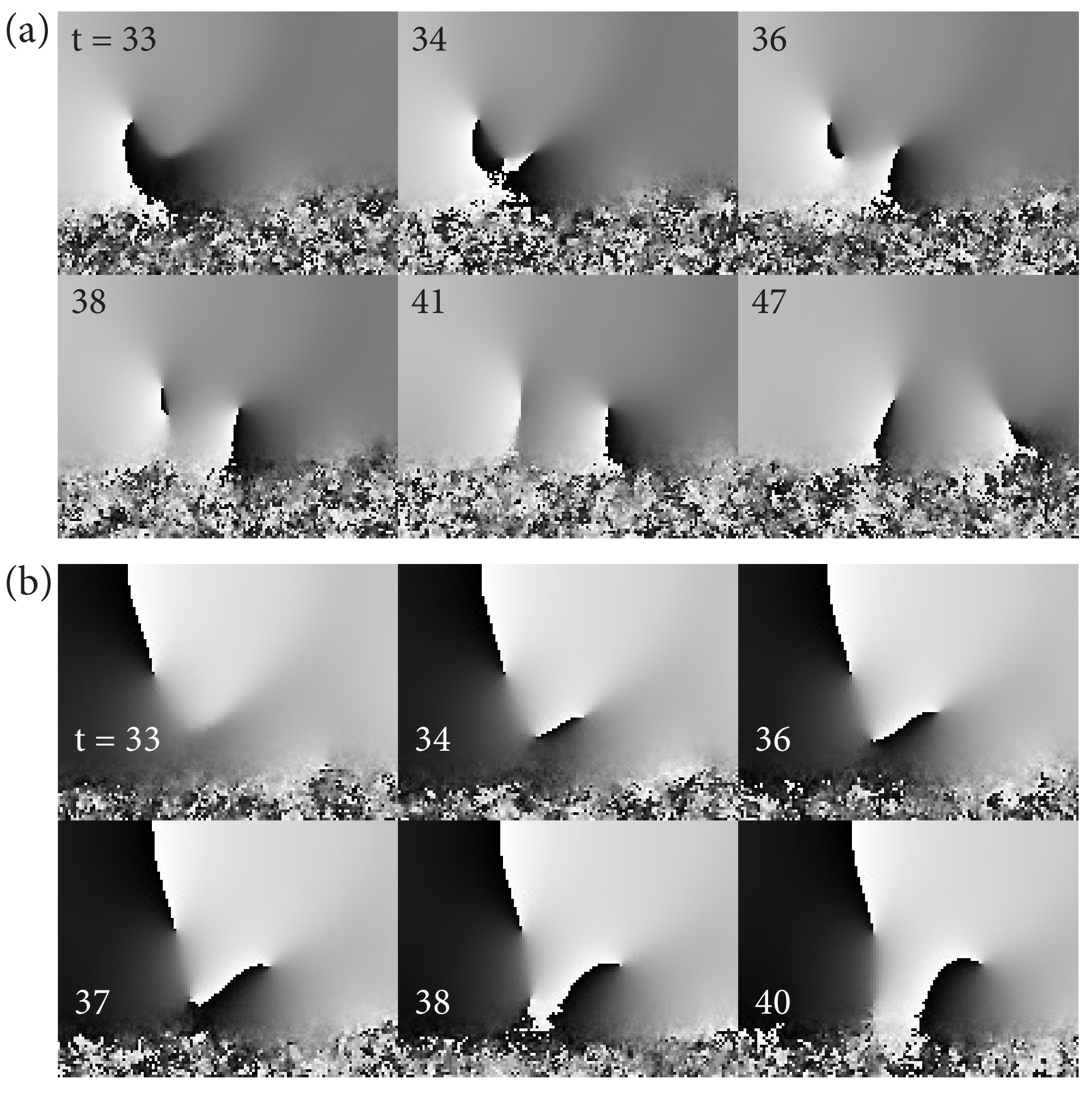}
\caption{Phase plots from $0$ (black) to $2\pi$ (white) of tearing
    processes in the 2D simulations of Fig.~\ref{fig:contours1} 
    with (a) a density cut perturbation and (b) a phase
imprinting perturbation, focused on the
evolution of the bottom-right vortex and phase wall structure. 
In the first 
frame at $t=33$ in both cases, 
a portion of the phase wall extends to the left and bottom. This strongly 
bent phase wall 
nucleates a new vortex-antivortex pair at $t=34$.
For (a), in the third and fourth frames ($t=36,\ 38$), 
the antivortex moves close to the original
vortex, annihilating and leaving a phase wall at $t=41$. Finally, at
$t=47$ the phase wall nucleates a new vortex, leaving the observed two
vortices described in the main text.
For (b) and in contrast to (a), 
in the third, fourth and fifth frames the antivortex
 moves toward the trap edge, and is eventually ejected in at $t=40$. This leads
to the same end result as (a), but without the vortex pair annihilation observed
in (a).
\label{fig:teardetail}}
\end{figure}

A key part of the tearing process is the elimination of a vortex/antivortex pair
on each side of the phase wall. Figure \ref{fig:teardetail} shows this process
in more detail, using the same simulations
as Fig.~\ref{fig:contours1}.
Figure \ref{fig:teardetail}(a) describes the tearing process from a density cut
perturbation in which the phase wall bends outward, forming a beveled shape
with its ends pinned by vortices.
The beveling nucleates a vortex-antivortex pair near each phase wall end. 
In each quadrant, the original vortex
annihilates with the antivortex of this new pair.
This leaves a phase wall which bends and forms a new vortex
resulting in two vortices in each quadrant.  Figure \ref{fig:teardetail}(b)
shows the same process from a 130$^{\circ}$
phase imprinting perturbation, but no pair annihilation occurs.  Instead, the
antivortex in the new pair is ejected. Both cases, however, 
result in the same two-vortex configuration in each quadrant, 
contributing to the generality
both of this tearing process and of the future cloud evolution.

\subsection{Vortex ejection and renucleation}
The final general, dynamical feature is
the repeated ejection and renucleation of at least some of the vortices. This
is shown in the final three frames of the time sequence displayed in 
Fig.~\ref{fig:contours2}.  We find that this process is driven by a specific
dynamic mechanism: when the
vortex pair forms, boundary effects cause the vortices to move rapidly toward
the ends of the trap, while the center of the phase gradient between the vortex
pair stays relatively stationary. This causes a ``back-bending'' of the phase
wall, which changes the orientation of the superfluid velocity near the
vortices, pointing partially out of the trap. This back-bending 
appears to eject
the vortices from the trap. Once the vortices have left, the phase wall 
bends forward and renucleates new vortices driven by the same
processes as described above.

\begin{figure*}
\includegraphics[scale=0.7]{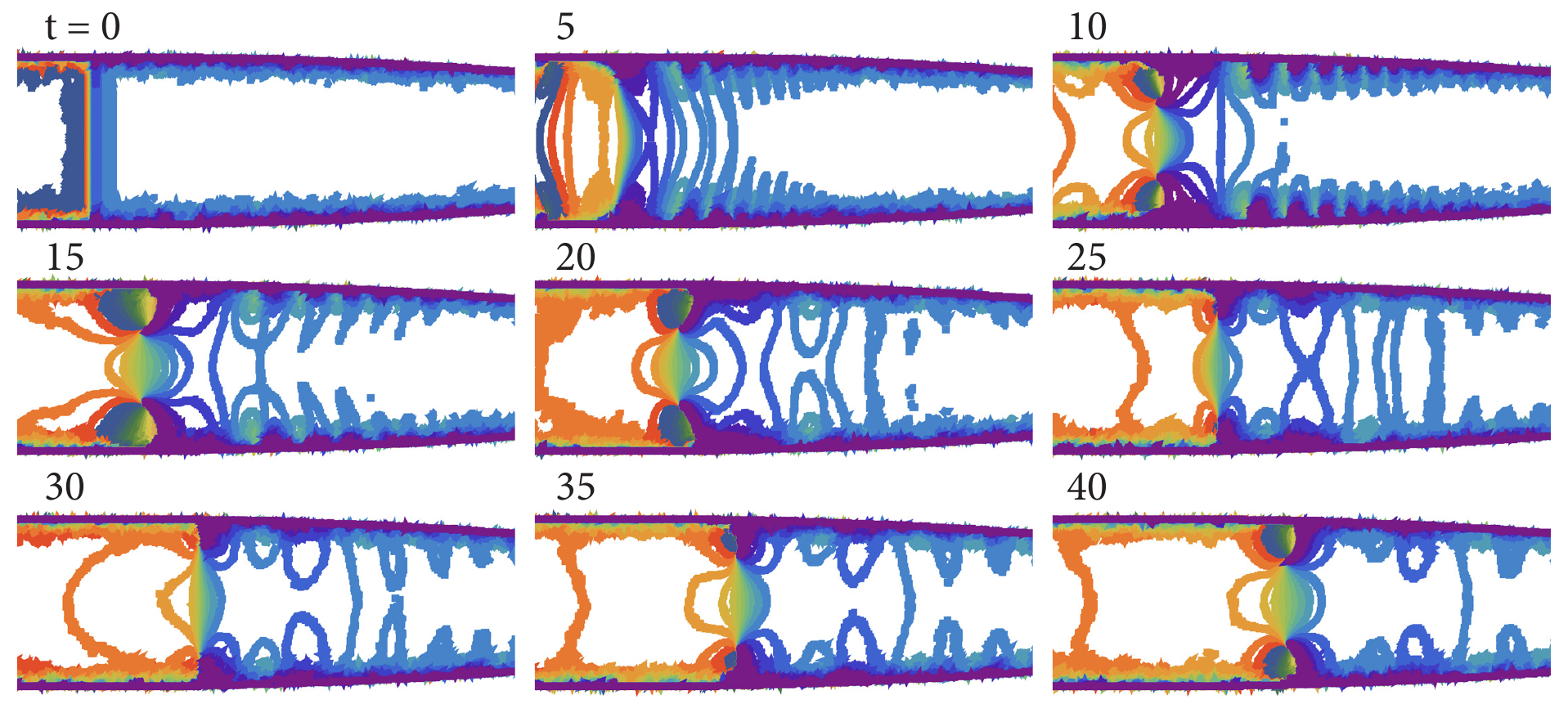}
\caption{Contour plots of the phase for a trap which is more tightly confined
radially ($\lambda = 13.2$).  In this regime, the phase imprinting perturbation
($t=0$) causes a phase wall to bend and nucleate a vortex pair ($t=10$), as in
the other simulations. In contrast to less-1D simulations, no ``tearing''
process can occur due to the additional confinement. Instead, the vortices
propagate until they are later ejected, again leaving a
phase wall ($t=25$). This leads to another sequence of phase wall bending and
subsequent vortex nucleation ($t=35$). A second
vortex ejection -- phase wall -- vortex nucleation sequence occurs but is not
shown here.
\label{fig:osc}}
\end{figure*}

\subsection{Summary of results}
The above steps are the general features
of these inhomogeneous, trapped
superfluids under two rather different perturbations. A few principles dictate
most of this behavior: trap inhomogeneity causes phase walls to bend; bent phase
walls nucleate vortex pairs and usually ``tear'' from the vortex pair; vortex
pairs can later be ejected due to back-bending of phase walls.
The differences between systems are differences in degree,
such as multiple tearing processes, or
vortex ejections and renucleations that differ in order and number.
The number of such events is most closely related to
the magnitude and steepness of the initial phase gradient induced
by the perturbation.

Beyond the first 
5 to 30 ms, in the later stages of equilibration, the
behavior becomes more unique to each system, again depending strongly on the
phase gradient applied. 
For large, sharp gradients the
defects reach the end of the trap, causing
a reflection that is dependent on the precise trap geometry, and can be much
more complex and varied compared to the 
steps reported above \footnote{In this
context, phase imprinting of $\Delta\Phi \approx
180^\circ$ reflects a special case of a strong gradient creating a defect with
very little velocity \cite{scherpelz_2014}. As described in our previous work
and Table \ref{tab:varypi} here,
the same initial sequence of phase wall bending and vortex ring nucleation
is observed, but ultimately, because of the small defect
velocity, an off-center ring results which
leads to a longer-lived precessing line vortex \cite{scherpelz_2014}.}.
For weaker
phase gradients the defects dissipate or exit the trap before reflecting
from the 
trap end, as explored further in App.~\ref{sec:vel}.  

\section{Trap dimensionality effects}

The above observations can be compared with earlier work, which investigated a
similar sequence in a closer-to-1D configuration. That work was
interpreted to be a ring--soliton oscillation
\cite{shomroni_2009}. We find that, instead, Ref.~\cite{shomroni_2009}
exhibits the same dynamic mechanism described above, but with small changes due
to the closer-to-1D configuration ($\gamma = 5.0$, $\lambda = 9$).  In
Ref.~\cite{shomroni_2009} the phase wall bends and nucleates a vortex ring as
described in this work.  However, the phase wall does not bend enough for part
of it to survive and tear away from the vortex ring, so only one defect
structure persists.  This vortex ring can then undergo the dynamic ejection and
renucleation process described above.

To make this comparison concrete, Figure \ref{fig:osc} 
shows phase contour plots for a simulation run that displays an apparent 
oscillation between phase wall and vortices, very similar to
Ref.~\cite{shomroni_2009}.
Here a $130^\circ$ phase imprinting
perturbation has been applied. These 2D simulations have the same
$\omega_z$ as other simulations presented in this work,
but a trap ratio $\lambda=\omega_\perp/\omega_z=13.2$, which moves the trap
closer to a one-dimensional geometry ($\gamma = 4.0$), quite similar to
Ref.~\cite{shomroni_2009}. This apparent oscillation
is present in both density cut and phase imprinting perturbations. 

\begin{figure}
    \includegraphics[scale=0.73]{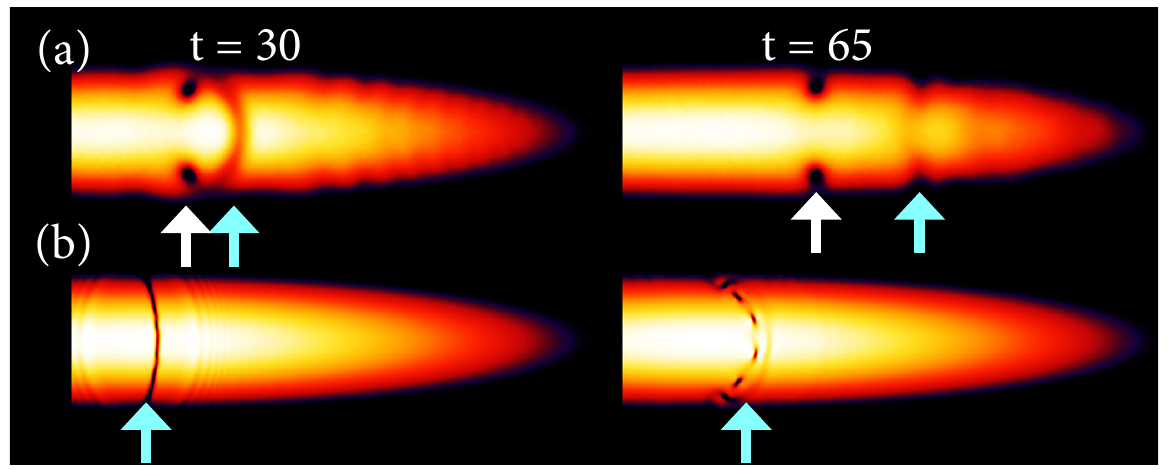}
    \caption{Plots comparing two different values of $\gamma$.  In (a), the
        amplitude of the wavefunction is displayed at two time points for the
        same trap parameters as in Fig.~\ref{fig:contours1}(b) ($\gamma =
        8$). In (b), a run with the same trap ratio $\lambda$, but much smaller
        $\omega_\perp$ and $\omega_z$, is shown at the same time points ($\gamma
        = 32$). The first frames, at $t = 30$, 
        demonstrate that the bending and ``tearing'' processes
        occur at very different timescales, due to the decreased local trap
        inhomogeneity. In the small trap, tearing is nearly complete; in the
        large trap, a vortex pair near the edge has not even formed yet. The
        large system is just beginning to tear at $t = 65$, while small system
        has a very well-separated vortex pair and phase wall. The $t=65$ frame
        in (b) also demonstrates the other result of large traps:
        vortex-antivortex pairs form in the middle of the phase wall that is
        present, which quickly lead to much more chaotic dynamics. In both of
        these runs, a $150^\circ$ phase imprinting was applied, which shows the
        vortex-antivortex creation more clearly than $130^\circ$ at this system
        size. Cyan arrows indicate the initial phase wall, white arrows a
        separated vortex pair.
    }
    \label{fig:largegamma}
\end{figure}

As this process follows the same
mechanism as that shown in earlier figures, which does not lead to oscillation, 
we view this as an extreme form of the dynamics described in the less-1D 
context presented above.  We do not consider it to be
a ``Rabi'' oscillation \cite{shomroni_2009}, which would imply an energetic
near-degeneracy that allows an oscillation. Instead, this process is a
consequence of non-equilibrium fluid
dynamics of the superfluid -- specifically the movement of vortices and phase
walls in the very inhomogeneous trapping potential, which causes the phase wall
to bend and the vortices to alternately be ejected and re-nucleated by the phase
wall.

In the other limit of trap dimensionality, for very three-dimensional traps
(large $\gamma$), the primary phenomena of the original phase wall bending
forward and nucleating a pair of vortices is quite general, as is tearing for a
wide range of $\gamma$ values. However, at very
large $\gamma$ two major factors enter, as displayed in
Fig.~\ref{fig:largegamma}. First, the trap inhomogeneity is small,
so the bending is much less pronounced, and tearing processes occur much later.
Second, the reduced trap inhomogeneity and large size give freedom for many
vortex-antivortex pairs to nucleate within the phase wall, 
making the system more chaotic. Especially for slow-moving phase walls
(phase imprinting angles closer to $180^\circ$), these pairs were found to 
play a role in the dynamics for $\gamma \gtrsim 20$ with the
parameters used here, with the largest
values of $\gamma$ producing the greatest number of pairs. These additional
vortices also resemble those of Fig.~5 of Ref.~\cite{dutton_2001}, where $\gamma
\sim 19$.

\section{Conclusion} 
In this paper, we
have demonstrated general features in the evolution of
planar defects in trapped superfluid gases. This common behavior
is demonstrated by establishing the similarity in
both phase imprinting and density
depletion. This work focuses on
the earliest stage of equilibration (at most about 30 ms), where
these two perturbations lead to planar defects of a very similar
fashion which are ultimately ``healed" during the
equilibration process.
Later stages involve distinct phenomena, not discussed here,
associated with
trap boundaries and sometimes ``solitonic" vortices
\cite{scherpelz_2014,ku_2014,wlazlowski_2014}.
Not only do a variety
of 
initial conditions or sudden
perturbations produce these sharp phase walls, but once
formed, they
lead to a predictable set of rather complex dynamical processes accompanying
superfluid equilibration.
This complexity, in turn, reflects the ability of
planar defects to form multiple topological 
features.

The steps in common involve first
a bending of the phase wall,  
next ``tearing'' that can create multiple vortex
rings and/or phase walls, and finally
vortex ejection and renucleation.
These all serve to heal planar defects, and the formation and loss of vortex
rings near the trap edge accompany and drive such healing processes.
We emphasize that
all of
these dynamical processes are consequences of trap inhomogeneity and 
boundary effects, in
contrast to other features, such as the snake instability, that may be seen in
more homogeneous systems.

Finally, these predictions resulting from our simulations 
should be accessible experimentally and thus directly testable. The complete
evolution of the planar defect, and especially the ``tearing'' process, should
be observable. In addition, performing both phase imprinting and density cut
perturbations would reveal the notable similarity in their effects. These
simulations can thus serve as a guide for understanding defect evolution in a
variety of future experiments.

This work is supported by NSF-MRSEC Grant
0820054. Work at Argonne was supported by 
the Scientific Discovery through Advanced Computing (SciDAC) program funded
by U.S.\ Department of Energy, Office of Science, Advanced Scientific Computing
Research 
and Basic Energy Sciences, Office of Science, Materials
Sciences and Engineering Division. 
 The numerical work was performed on NIU's
GPU cluster GAEA.
Finally, we are grateful to William Irvine and Adam Ran{\c{c}}on
for insightful discussions.

\appendix

\begin{figure}
\includegraphics[scale=0.65]{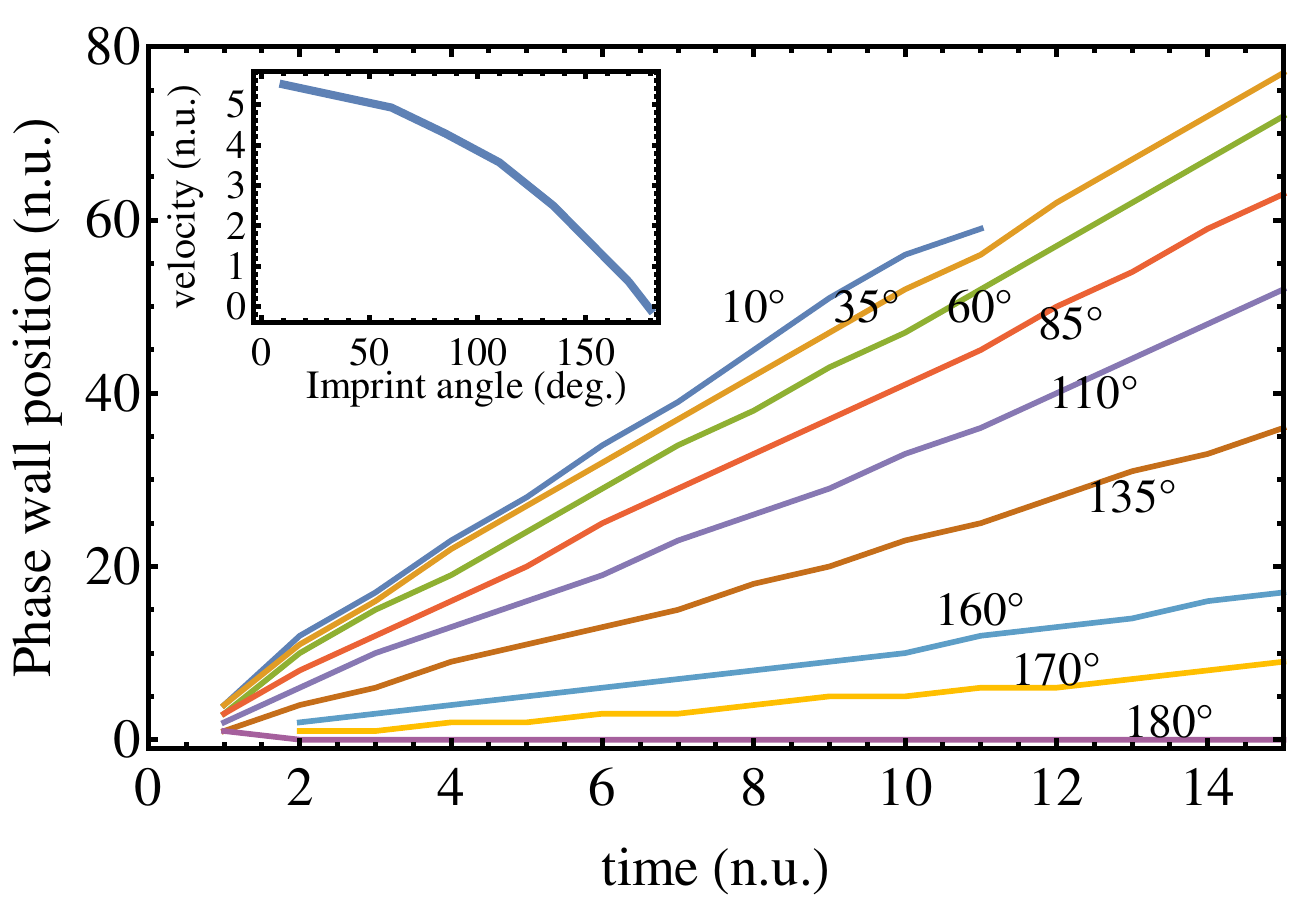}
\caption{Plots of the position of the phase wall as a function of time,
depending on the phase imprinting angle used, in terms of normalized units
(n.u.). The phase wall position is
calculated as the maximum magnitude of the phase gradient $\partial \phi/
\partial z$ along the central $z$-axis,
measuring from the trap center $z=0$. The inset shows the calculated average
velocity over this timescale, as a function of phase imprinting angle.
Over this time scale, the phase imprinting
angle dramatically changes the phase wall velocity. Only at the extremes of
angles (see Table \ref{tab:varypi} for details), however, do the qualitative
dynamics vary.
\label{fig:phasevel}}
\end{figure}

\begin{table}
    \begin{tabularx}{\linewidth}{|l|XXXXX|}
        \hline
        Angle & Vortices & Tearing & 2nd Tearing & Re-nucleation &
         Tilting \\
        \hline
        $10^\circ$  &   &   &   &   &   \\
        $35^\circ$  &   &   &   &   &   \\ 
        $60^\circ$  & Y &   &   & U &   \\
        $85^\circ$  & Y & U &   & Y &   \\
        $110^\circ$ & Y & Y &   & Y &   \\
        $135^\circ$ & Y & Y & Y & Y &   \\
        $160^\circ$ & Y & Y &   & Y & Y \\
        $170^\circ$ & Y & Y &   & Y & Y \\
        $180^\circ$ & Y & N/A & N/A & N/A & N/A \\
        \hline
    \end{tabularx}
    \caption{Table of phenomena seen in defect evolution as the phase imprinting
    angle is varied. ``Y'' means the phenomenon is observed. ``U'' means it is
    uncertain, usually because the dynamics occur in a region of extremely low
    density. A traveling phase wall forms initially for all runs between
    $10^\circ$ and $170^\circ$. For $180^\circ$, there is no traveling phase
    wall, and multiple vortices nucleate immediately. As a result, the system
    enters a much more chaotic state, and further phenomena cannot be 
    identified. Here ``tearing'' is defined as a process in which one phase wall
visibly separates into two phase walls; ``2nd tearing'' is seeing one of these
phase walls again split into two.  ``Re-nucleation'' means vortices are observed
exiting the trap, and new vortices are later nucleated at a similar position.}
    \label{tab:varypi}
\end{table}

\section{Videos}
\begin{figure}
\includegraphics[scale=0.36]{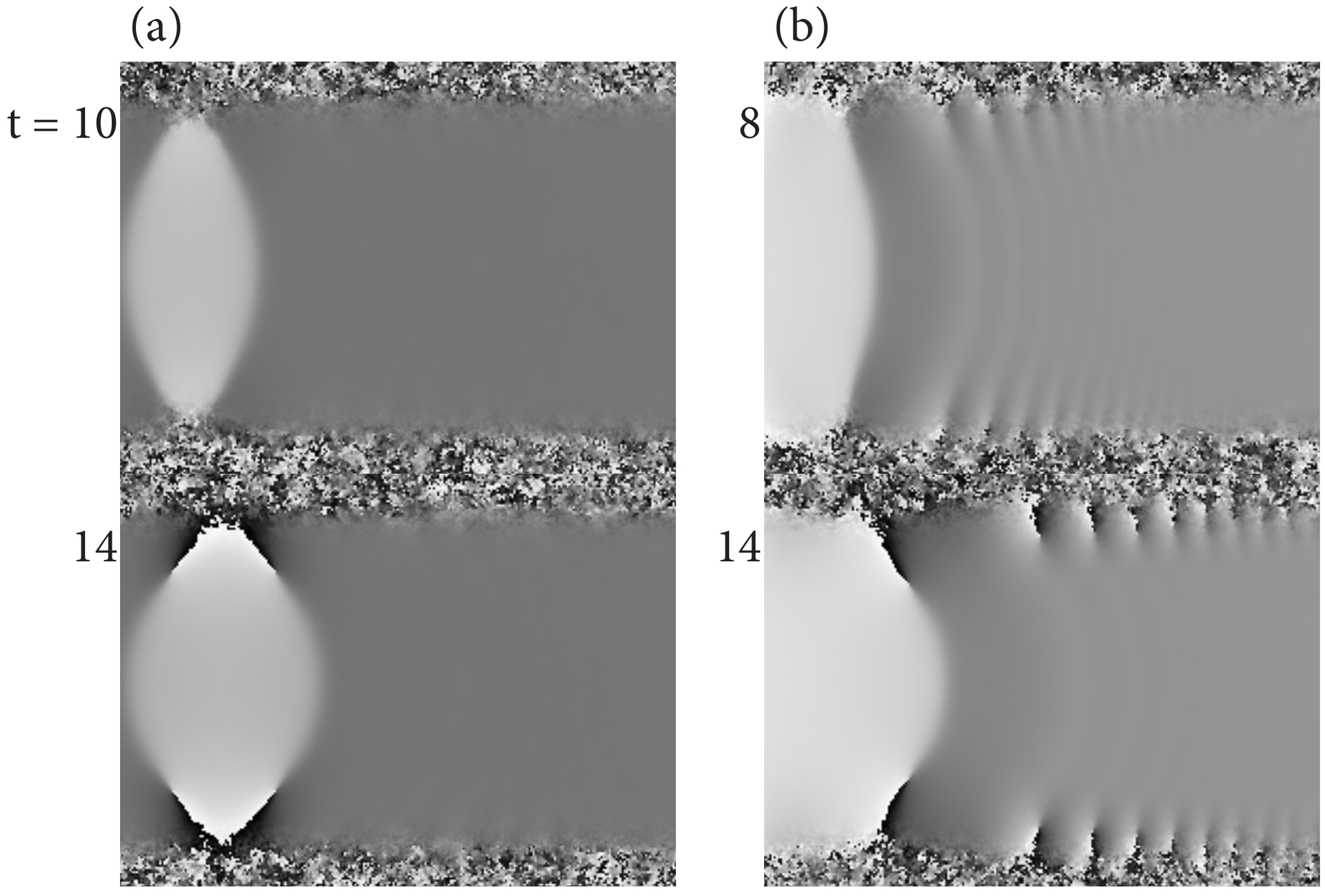}
\caption{Phase plots of (a) density cut and (b) 130$^{\circ}$ phase imprinting
perturbations in an inhomogeneous, 2D cylindrical trap. Both simulations use
a $z$-axis square well potential of inner length $\Delta z = 160$, and a
radial trapping frequency of $\omega_\perp = 0.12\mu/\hbar$, giving a similar
overall trap shape to the cigar trap in the main text.  
Shown here is the initial vortex pair creation resulting from
the perturbation in both cases. For both perturbations, and 
just as in the case of the cigar trap, a
phase wall forms and bends ($t=10$ or $8$), and the bending soon nucleates vortices
that move inward ($t=14$).
\label{fig:VCCyl}}
\end{figure}

\begin{figure}
\includegraphics[scale=0.36]{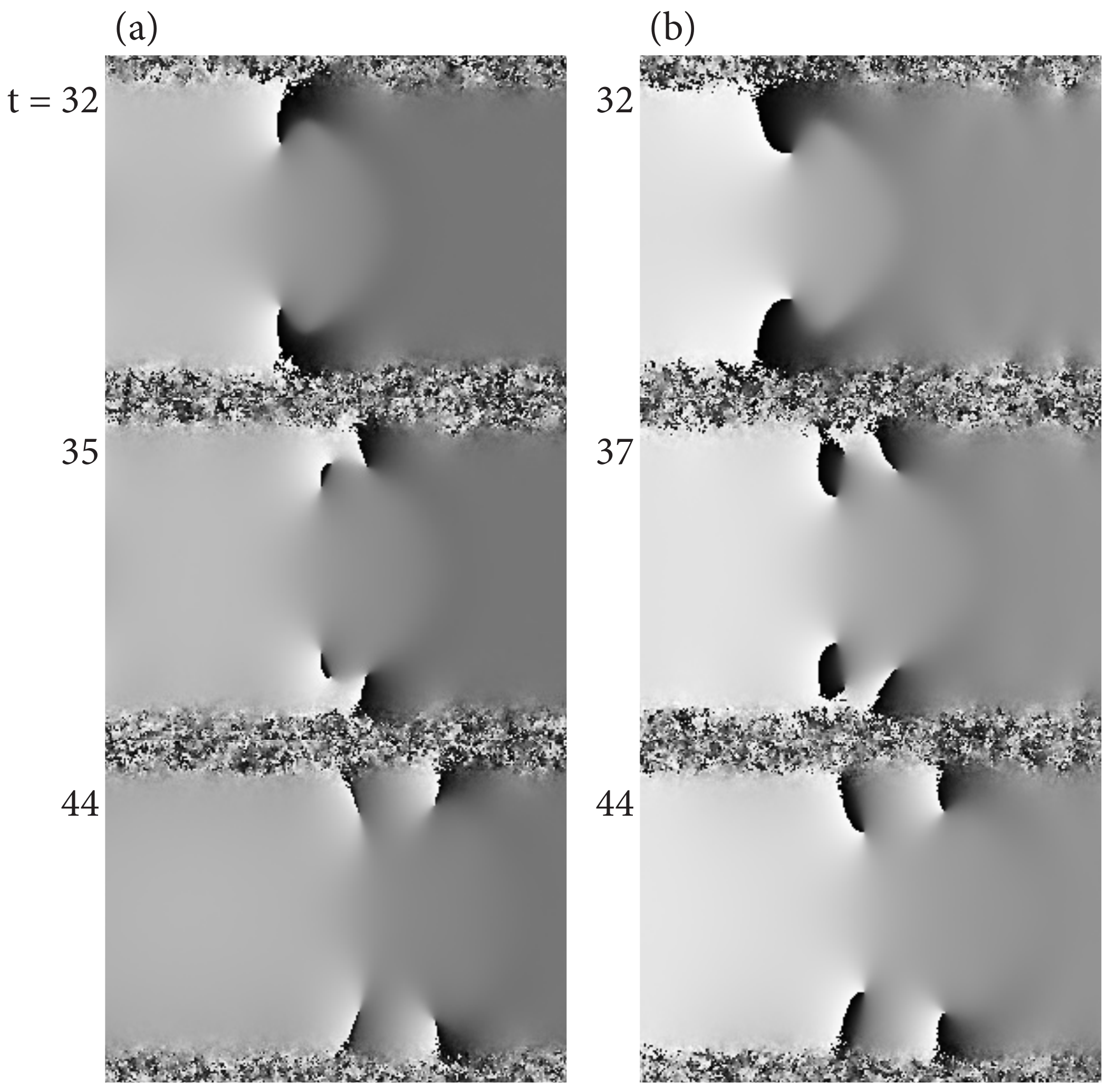}
\caption{Phase plots of a cylindrical trap for the same runs as Figure
\ref{fig:VCCyl}. Shown here is the tearing process that follows vortex pair
nucleation. In the first frames of both (a) and (b), a phase wall bends outwards
from the vortex pair that formed earlier. In the second frames, a
vortex-antivortex pair is created on both the top and bottom of the system, and
the antivortices proceed to (a) 
annihilate with the original vortices present in the
system, or (b) rapidly exit the system. 
Finally, in the third frames two vortices of the same circulation 
are present along each edge. Again, these phenomena 
agree both between density cuts and
phase imprinting, and between inhomogeneous cylindrical and cigar-shaped traps.
\label{fig:TCyl}}
\end{figure}

\begin{figure}
\includegraphics[scale=0.36]{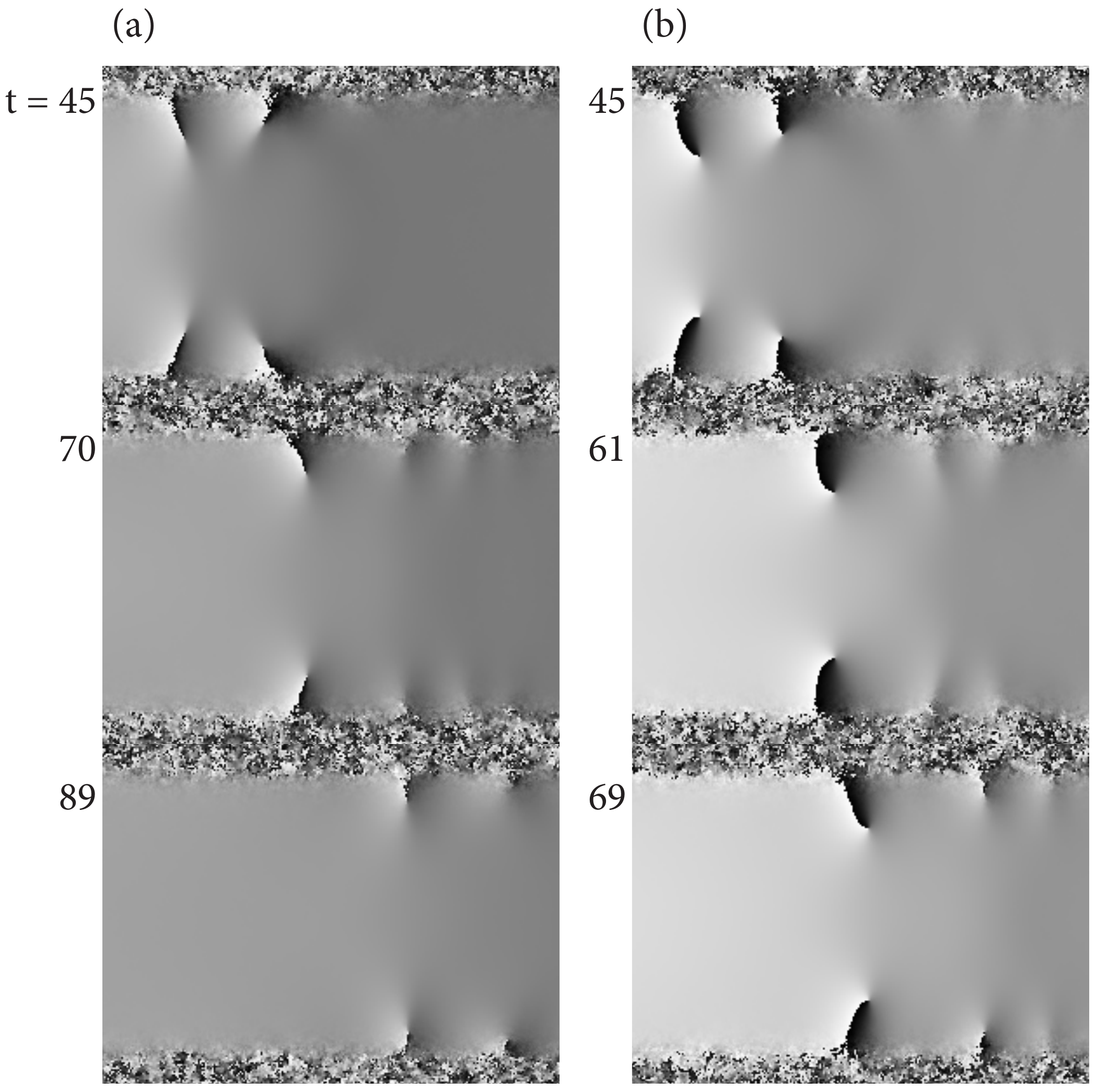}
\caption{Phase plots of a cylindrical trap for the same runs as Figure
\ref{fig:VCCyl}. These frames display the vortex exit and renucleation. In both
(a) and (b), 
the front vortices are ejected in the second frame ($t=70$ or $61$), and
renucleate,
albeit very close to the boundary, in the third frame ($t=89$ or $69$). 
As before, these dynamics mirror those in cigar-shaped traps. Note that compared
to Fig.~\ref{fig:VCCyl} and Fig.~\ref{fig:TCyl}, the section of the cloud
shown here is shifted to the right.
\label{fig:VERCyl}}
\end{figure}

Available in the Supplemental Material \cite{defects_sm_2014}  are a
selection of videos for different simulation 
runs. ``Cut3D'' and ``PI3D'' refer to the 3D density
depletion and phase imprinting simulations of Fig.~\ref{fig:tearing}. 
``Cut2D'' and ``PI2D'' show the runs in Fig.~\ref{fig:contours1}(a) and
Fig.~\ref{fig:contours1}(b) of the main text, respectively. Finally, 
``Cyl'' shows the density cut simulation in Figs.~\ref{fig:VCCyl}(a) 
through \ref{fig:VERCyl}(a).

All videos show the phase plot from 0 (black) 
to $2\pi$ (white) on the top half of the frame, and on the bottom half show
the density (for 3D, density in the $x=0$ plane), scaled from 0
(black) to a run-dependent maximum density (white).

\section{Variations of the phase imprinting angle\label{sec:vel}}
Figure \ref{fig:phasevel} demonstrates the effect of changing the phase
imprinting angle $\Delta\Phi$ that is applied in phase imprinting-type 
perturbations. In
all cases, we find that the phase wall bends and begins to move outward. As
noted in our earlier work \cite{scherpelz_2014}, 
the phase imprinting angle does influence 
whether a vortex line will form,
but it has little effect on the lifetime of the vortex ring. Here,
Fig.~\ref{fig:phasevel} shows that the phase imprinting angle changes the
velocity of the center of the phase wall. Qualitative differences only arise due
to limits of the phase angle applied, as shown in Table \ref{tab:varypi}. At
very small $\Delta\Phi < 60^\circ$, 
not enough of a perturbation is applied to form
vortices at all.  At $\Delta\Phi \geq 160^\circ$, the velocity of the phase wall
is very small, so that the vortex pair tilts and one vortex is ejected. This
situation was analyzed in Ref.~\cite{scherpelz_2014}. Between these two limits,
the behavior is quite general.

\section{Cylindrical traps\label{sec:cyl}}

Figures \ref{fig:VCCyl} through
\ref{fig:VERCyl}  characterize the defect sequence in an inhomogeneous
cylindrical geometry. These defect sequences match very well with that
described in the main text, which is due to the dominant influence of the radial
inhomogeneity present in both systems. In these cylindrical traps, a strong
square-well potential is applied along the $z$-axis, while a harmonic trapping
potential is used in the radial direction. Figures \ref{fig:VCCyl} through
\ref{fig:VERCyl} display three characteristic parts of the defect sequence for
both density cuts and phase imprinting, all of which match closely with the
defect sequences described in the main text.

By contrast, for a ``hard-walled'' cylindrical trap, in which the potential in
the $x-y$ plane is $0$ inside a critical radius and very large outside that
radius, the behavior is very different. This is due to the lack of inhomogeneity
in the superfluid density. 
Instead, we observe only two defects: an extremely short-lived, small-radius 
vortex ring which nucleates then annihilates in the
center, and vortices that nucleate right at the boundary but do not move inward
or otherwise affect phase wall propagation. Both of these defects seem to 
carry very little energy, and have negligible effects on the dynamics. Other
than these defects, no interesting evolution is observed -- the phase walls
simply travel towards the trap edges and slowly dissipate. When the phase walls
 reach the end, no defects remain (no reflections are present).

\bibliography{ReviewArxiv}

\end{document}